\documentclass[a4paper]{PoS}
\pdfoutput=1

\usepackage{amsmath}
\usepackage{booktabs}
\usepackage{microtype}

\newcommand{\vecd}{\mathbf{d}}
\newcommand{\vecP}{\mathbf{P}}

\newcommand{\Ktilde}{\widetilde{K}}

\newcommand{\Ktildeinv}{\widetilde{K}^{-1}}

\newcommand{\qcm}{\mathbf{q}_{\rm cm}}
\newcommand{\qq}{\qcm^{2}}
\newcommand{\Ecm}{E_{\rm cm}}

\newcommand{\bra}[1]{\langle#1|}
\newcommand{\ket}[1]{|#1\rangle}
\newcommand{\expect}[1]{\langle#1\rangle}

\title{$K\pi$ scattering and excited meson spectroscopy using the stochastic LapH method}

\ShortTitle{$K\pi$ scattering and meson spectroscopy using stochastic LapH}

\author{\speaker{Ruair\'i Brett}$\;^a$, John Bulava$^b$, Jacob Fallica$^c$,
  Andrew Hanlon$^d$, Ben H\"orz$^e$, Colin Morningstar$^a$\\
  \llap{$^a$}Dept. of Physics, Carnegie Mellon University,
  Pittsburgh, PA 15213, USA\\
  \llap{$^b$}Dept. of Mathematics and Computer Science and CP3-Origins,
  University of Southern Denmark,
  Campusvej 55, 5230 Odense M, Denmark\\
  \llap{$^c$}Department of Physics and Astronomy, University of Kentucky,
  Lexington, KY 40506, USA\\
  \llap{$^d$}Helmholtz-Institut Mainz, Johannes Gutenberg-Universit\"at,
  55099 Mainz, Germany\\
  \llap{$^e$}PRISMA Cluster of Excellence and Institute for Nuclear Physics,
  Johannes Gutenberg-Universit\"at,
  55099 Mainz, Germany\\
  E-mail: \email{rbrett@cmu.edu}}

\abstract{Elastic $I=1/2$, $s$- and $p$-wave $K\pi$ scattering amplitudes are
  simultaneously calculated using a L\"uscher style analysis on a single ensemble of
  dynamical Wilson-clover fermions at $m_\pi \sim 230$~MeV. Partial wave mixing due to
  the reduced rotational symmetries of the finite volume is included up to $\ell=2$.
  We also present finite-volume QCD spectra on two large anisotropic lattices
  ($32^3 \times 256$, $24^3 \times 128$) with $m_\pi \sim 230,\ 390$ MeV respectively.
  In each symmetry channel, a large basis of one- and two-hadron interpolating operators
  is employed with all-to-all quark propagation treated using the stochastic LapH method.}

\FullConference{The 36th Annual International Symposium on Lattice Field Theory
  - LATTICE2018\\ 22-28 July, 2018\\
  Michigan State University, East Lansing, Michigan, USA.}

\begin{document}
\section{Introduction}
As most excited hadrons appear as unstable resonances in experimental scattering
cross sections, to study such states, first-principles determinations of hadron-hadron
scattering amplitudes in QCD are desirable.  Monte Carlo estimates in lattice QCD must
be done in finite volume and using imaginary time, so directly determining scattering
amplitudes is not possible.  Using a particularly
successful approach introduced by L\"uscher, this difficulty can be circumvented by
inferring infinite-volume scattering amplitudes from interacting finite-volume
energies~\cite{Luscher1991}. Further developed in
Refs.~\cite{Rummukainen1995,Kim2005,Gockeler2012,Briceno2013a,Briceno2014,Morningstar2017},
among others, the method is now well established in the calculation of 2-to-2 scattering
amplitudes. In this talk, a recent determination~\cite{Brett2018a} of elastic $K\pi$
scattering amplitudes is presented, where the partial wave mixing due the the
finite-volume is treated explicitly. Using the procedure outlined in
Ref.~\cite{Morningstar2017}, $s$- and $p$-wave
amplitudes are determined, the latter being well-described by a Breit-Wigner form, as
expected in the presence of a narrow $K^*(892)$ resonance.

Above three-particle thresholds, the formalism for extracting scattering observables
is maturing (e.g., Refs.~\cite{Briceno2017b,Doring2018}), with the first application to
QCD appearing recently in Ref.~\cite{Mai:2018djl}. As such, understanding
the resonant spectrum in this regime from first principles calculations remains a
challenge. The second half of this talk presents preliminary results from a qualitative
determination of the excited spectrum of QCD in a set of $I=1$ bosonic symmetry channels.
By employing large bases of interpolating fields for single- and multi-hadron operators
both the single-hadron dominated states and states with significant mixing with
multi-hadron operators can be disentangled.

\section{Finite-volume spectrum determination}
In order to extract the finite-volume spectrum in a given symmetry channel, the
$N \times N$ matrix $C_{ij}(t) \equiv \expect{O_i(t+t_0) \overline{O}_j(t_0)}$ of
correlation functions is evaluated using the stochastic LapH
method~\cite{Morningstar2011}. By solving the generalised eigenvalue problem (GEVP)
\begin{equation} \label{eq:gevp}
  C(t_d)v(t_0,t_d) = \lambda(t_0,t_d)C(t_0)v(t_0,t_d),
\end{equation}
for a single pair of diagonalisation times $(t_0,t_d)$, the diagonal
elements of the `rotated' correlation matrix formed by the eigenvectors ${v_n(t_0,t_d)}$
\begin{equation} \label{eq:rotated_corr}
  \widetilde{C}_n(t) = (v_n,C(t)v_n),
\end{equation}
can be fit with single or multi-exponential fits to extract the $N$ lowest energies in the
spectrum. The overlaps $Z_{j}^n = \bra{0}O_j\ket{n}$ between the initial set of
operators and the finite-volume eigenstates can then be estimated.
For the scattering amplitude analysis described in Sec.~\ref{sec:luscher}, as the
signal of interest is the deviation of two-particle energies from their non-interacting
counterparts, the so-called \textit{ratio fits} of Ref.~\cite{Bulava2016} are used.

%% As discussed
%% in sec.~\ref{sec:luscher}, for the estimation of scattering amplitudes the signal of
%% interest is the deviation of finite-volume two-hadron energies from their non-interacting
%% counterparts. To this end, the energy difference $\Delta E$ is extracted from a
%% single-exponential fit to the ratio
%% \begin{equation} \label{eq:ratio_fits}
%%   R_n(t) = \frac{\widetilde{C}_c(t)}{C_\pi(\vecd^2_\pi,t)C_K(\vecd^2_K,t)},
%% \end{equation}
%% where the nearest non-interacting state to level $n$ consists a pion and kaon with momenta
%% $\frac{2\pi}{L}\vecd_\pi$ and $\frac{2\pi}{L}\vecd_K$ respectively.

\section{Scattering amplitudes from finite-volume energies}
\label{sec:luscher}
Finite-volume energies are determined in the `lab' frame in which the two-particle
systems may have non-zero total momentum. In the centre-of-mass frame we define for $K\pi$
scattering the following kinematic quantities
\begin{equation}
  \Ecm = \sqrt{E^2-\vecP^2_{\rm tot}}, \qquad
  \qq = \frac{1}{4}\Ecm^2 - \frac{1}{2}(m_\pi^2+m_K^2)
  + \frac{(m_\pi^2-m_K^2)^2}{4\Ecm^2},
\end{equation}
where $E$ is the lab frame energy determined above.
The relationship between the two-particle centre-of-mass energies and the infinite-volume
scattering amplitude can be expressed as~\cite{Morningstar2017}
\begin{equation} \label{eq:quantisation_condition}
  \det[\Ktildeinv(\Ecm) - B^{(\Lambda,\vecd)}(\Ecm)] = 0,
\end{equation}
which holds up to exponentially suppressed corrections in the spatial extent $L$.
$\Ktildeinv$ and $B$ are infinite-dimensional matrices in partial wave $\ell$, and are
real and symmetric, and Hermitian respectively, ensuring the determinant itself is real
for real $\qq$.
Expressions and software for numerical evaluation of the $B$-matrix
elements are provided in Ref.~\cite{Morningstar2017}. For the scattering of spinless
particles, following the convention of Ref.~\cite{Brett2018a},
%% For unitary elastic scattering matrix $S$, $K$ is real, symmetric, and diagonal in $\ell$
%% and $n_{\rm occ}$. They are related by
%% \begin{equation} \label{eq:Kmatrix}
%%   K = (2T^{-1} + i)^{-1},\qquad S = 1 + iT,
%% \end{equation}
%% while for scattering of spinless particles, following the convention of~\cite{Brett2018a},
\begin{equation}
  \Ktildeinv_{\ell}(\Ecm) \equiv
  \left(\frac{\qcm}{m_\pi}\right)^{2\ell+1} K^{-1}_\ell(\Ecm)
  = \left(\frac{\qcm}{m_\pi}\right)^{2\ell+1} \cot \delta_\ell (\Ecm),
\end{equation}
is expected to be smooth near the elastic threshold. In the determinant
condition in Eq.~(\ref{eq:quantisation_condition}), partial wave mixing due to the reduced
symmetry of the cubic finite-volume must be treated with care. In order to proceed
with a practical computation, some truncation in $\ell$ is required. For a one-dimensional
$B$, the determinant condition is, of course, trivial, yielding a one-to-one relationship
between a finite-volume energy $\Ecm$ and an amplitude point $\Ktildeinv(\Ecm)$. Here,
however, as we consider $\ell \leq 2$, a parametrisation of each partial wave in
$\Ktildeinv$ is required with some number of fit parameters.

Irreps of the appropriate little group for various total momenta used in the $K\pi$
scattering analysis are listed in Table~\ref{tab:irreps}. While there are a number of
irreps in which the $\ell=1$ partial wave can be isolated, it is only the $A_{1g}$ irrep
at zero total momentum where $\ell=0$ amplitude points can be unambiguously obtained.
Hence, we determine both amplitudes simultaneously in the elastic region using the
\textit{determinant residual} method of Ref.~\cite{Morningstar2017}.

\begin{table}[!ht]
  \centering
  \begin{tabular}[t]{c c l}
    \toprule
    $\vecd$ & \textbf{$\Lambda$} & $\ell$ \\
    \midrule
    $(0,0,0)$ &$A_{1g}$& 0, 4, \ldots \\
    &$T_{1u}$& 1, 3, \ldots \\
    \midrule%
    $(0,0,n)$ &$A_1$& 0, 1, 2, \ldots \\
    &$E$& 1, 2, 3, \ldots \\
    \bottomrule
  \end{tabular}
  \hspace{3em}
  \begin{tabular}[t]{c c l}
    \toprule
    $\vecd$ & \textbf{$\Lambda$} & $\ell$ \\
    \midrule%
    $(0,n,n)$ &$A_1$& 0, 1, 2, \ldots \\
    &$B_1$& 1, 2, 3, \ldots \\
    &$B_2$& 1, 2, 3, \ldots \\
    \midrule%
    $(n,n,n)$ &$A_1$& 0, 1, 2, \ldots \\
    &$E$& 1, 2, 3, \ldots \\
    \bottomrule
  \end{tabular}
  \caption{Irreps $\Lambda$ of the appropriate little group for various
    total momenta $\vecP_{\rm tot} = (2\pi/L)\vecd$ (where $\vecd$ is a vector
    of integers) considered in this work. We consider $K\pi$ systems at rest as well
    as those with non-zero total on-axis, planar-diagonal, and cubic-diagonal momenta.
    These momentum classes are listed in the first column, where $n\in \mathbb{Z}$.}
  \label{tab:irreps}
\end{table}

\section{Results: $K\pi$ scattering amplitudes}
\begin{figure}[!ht]
  \centering
  \includegraphics[width=0.475\textwidth]{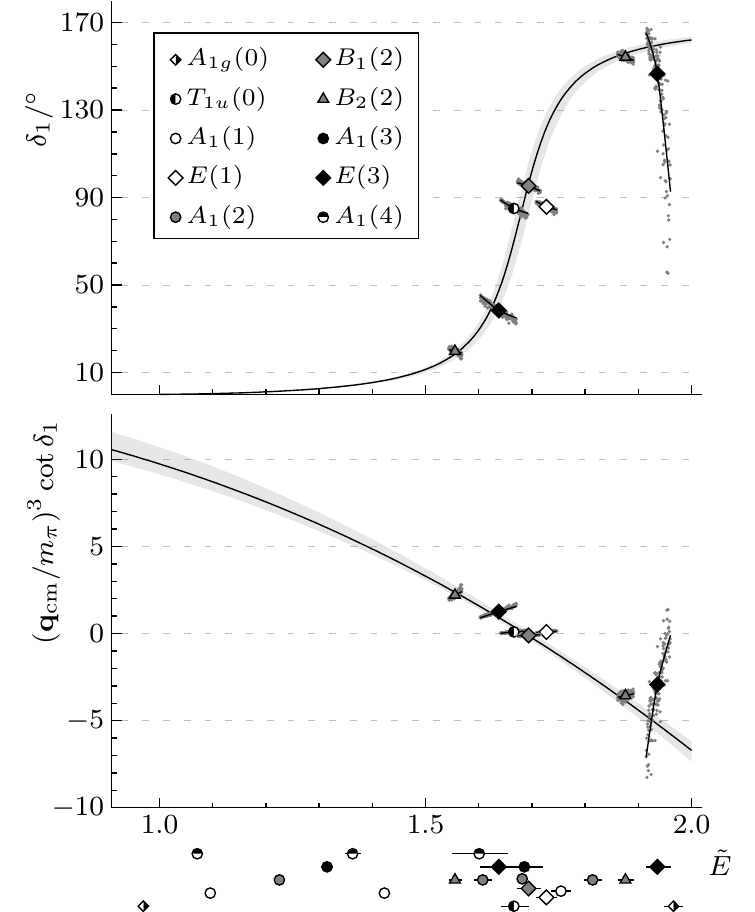}
  %% \hspace{-1.2em}
  \includegraphics[width=0.5\textwidth]{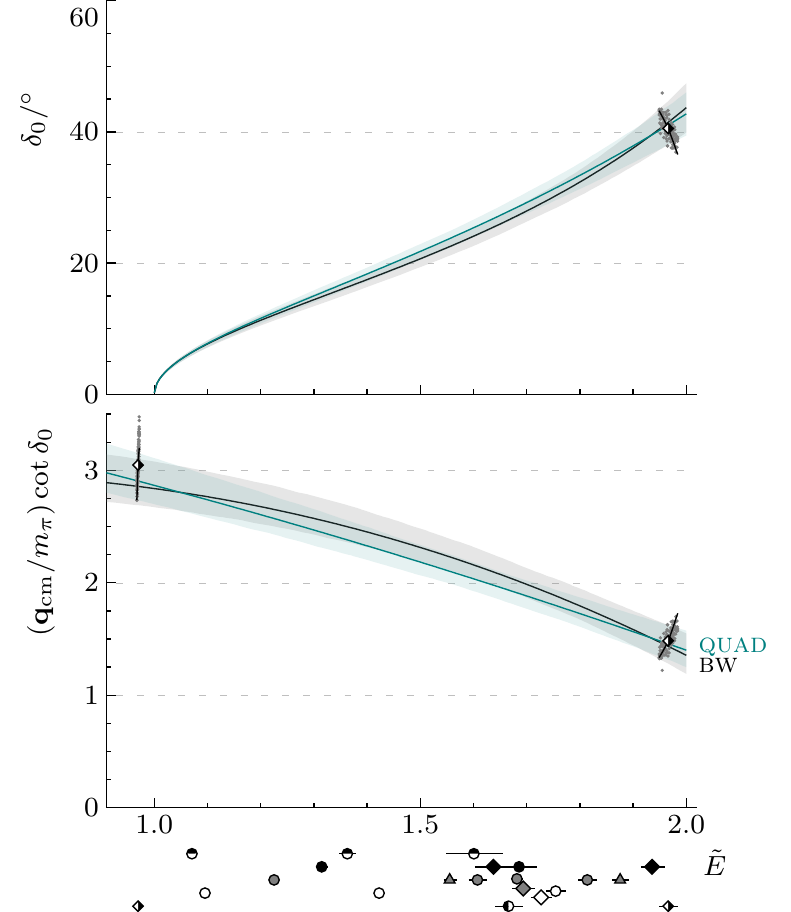}
  \caption{$K$-matrix fits to the $s$- and $p$-wave amplitudes as a function of
    $\tilde{E} = (\Ecm - m_K)/m_\pi$ such that the elastic region of interest extends over
    $1<\tilde{E}<2$. Together with the fits, which are explained in the text, we show
    amplitude points (neglecting $d$-wave contributions) from irreps with no mixing
    between $s$- and $p$-wave. All energies involved in the fit are indicated below the
    plots where they are offset vertically for clarity.
  \label{fig:cot}}
\end{figure}

Based on the expectation of a narrow $K^*(892)$ resonance, the $p$-wave amplitude is well
parametrised by a relativistic Breit-Wigner. For the $s$-wave amplitude we employ a
variety of fit forms including generic linear and quadratic functions of $\Ecm$,
motivated by analyticity of $\Ktildeinv(\Ecm)$ at threshold in $\Ecm$ and $s=\Ecm^2$
respectively. Additionally, we consider the NLO effective range expansion and an $s$-wave
relativistic Breit-Wigner to parametrise the amplitude. As a test of the validity of
the truncation in $\ell$, the $\Ktilde$- and $B$-matrices are also enlarged to include
$d$-wave contributions that we parametrise with the leading-order effective range
expansion. Explicit expressions are provided in Ref.~\cite{Brett2018a}.
Simultaneous fit results from a single anisotropic ensemble with $N_f = 2+1$
clover-improved Wilson fermions, $(32^3|230)$, with $m_\pi \sim 230$ MeV are listed
in Table~\ref{tab:fit}. It is apparent that the $K^*(892)$ resonance parameters are
insensitive to $s$-wave parametrisation and the inclusion of $d$-wave mixing. The
amplitudes from fit 2 are shown in Fig.~\ref{fig:cot}, together with the Breit-Wigner
$s$-wave amplitude from fit 5, illustrating that different parametrisations for the
$s$-wave produce a similar energy dependence in the elastic region. In addition to the
fits, points from irreps without $\ell=0,1$ partial wave mixing are shown and seen to
be consistent with the fit ans\"{a}tze.

In the energy range in which we have determined the $s$-wave amplitude, some hint of
the $K_{0}^{*}(800)$ may be expected to appear. From the LO effective range expansion,
$m_{\pi}a_0 < 0$ suggests a virtual bound state. However, as the ratio $1-2r_0/a_0$ must
be positive in the presence of a (real or virtual) bound state, using the NLO effective
range parameters from fit 3 gives $1-2r_0/a_0 = -8.9(2.4)$. At the $3-4\sigma$ level
then, we do not see any near-threshold bound state.

A careful analytic continuation, presumably requiring a better energy resolution than we
have here, is needed to establish the existence of a $K_{0}^{*}(800)$ resonance pole
above threshold on the second (unphysical) Riemann sheet.
Nevertheless, we can obtain qualitative information about a possible $s$-wave pole
by finding the zeros of $\qcm\cot \delta_0 -i\qcm$. This is easily done using the NLO
effective range parametrization of fit 3 and solving the resultant quadratic polynomial,
yielding $m_{R}/m_{\pi} = 4.66(13) -0.87(18)i$ which is consistent with the Breit-Wigner
mass and width from fit 4, which gives $m_{K_{0}^{*}}/m_{\pi} = 4.59(11)$ and
$g_{K_0^{*}K\pi} = 3.35(17)$. It is important to remember here that in
addition to the $K_{0}^{*}(800)$, the $s$-wave amplitude may also be
influenced by the $K_0^{*}(1430)$ resonance. Without a full analytic continuation
we can only infer qualitative information about a possible $s$-wave resonance pole from
the elastic $\ell=0$ amplitude calculated here.

\begin{table}[!ht]
  \centering
  \begin{tabular}{c c c c c c}
    \toprule
    Fit & $s$-wave par. & $m_{K^*}/m_{\pi}$ & $g_{K^*K\pi}$ &
    $m_{\pi}a_0$ & $\chi^2/\mathrm{d.o.f.}$ \\
    \midrule
    1 & \textsc{lin} & $3.810(18)$ & $5.30(19)$ & $-0.349(25)$ & $1.49$ \\
    2 & \textsc{quad} & $3.810(18)$ & $5.31(19)$ & $-0.350(25)$ & $1.47$ \\
    3 & \textsc{ere} & $3.809(17)$ & $5.31(20)$ & $-0.351(24)$ & $1.47$ \\
    4 & \textsc{bw} & $3.808(18)$ & $5.33(20)$ & $-0.353(25)$ & $1.42$ \\
    5 & \textsc{bw} & $3.810(17)$ & $5.33(20)$ & $-0.354(25)$ & $1.50$ \\
    \bottomrule
  \end{tabular}
  \caption{Results for the $K^*(892)$ resonance parameters and the
    $s$-wave scattering length $m_{\pi}a_0$ from all fits to the amplitudes.
    For each fit, the $p$-wave amplitude is described using a relativistic Breit-Wigner.
    Fit 5 includes $d$-wave contributions as discussed in the text.}
  \label{tab:fit}
\end{table}

%% \begin{figure}[!ht]
%%   \centering
%%   \includegraphics{figures/box_plot_no_overlaps.tikz}
%%   \caption{All finite-volume two-hadron energies boosted to the
%%     center-of-mass frame determined from single-exponential fits. Each irrep is located
%%     in one column, where the energies are shown in the upper panel as boxes with a
%%     vertical dimension equal to the statistical error, the non-interacting two-hadron
%%     levels as solid horizontal lines, and the relevant thresholds as dashed gray lines.}
%%   \label{fig:box}
%%   %% The corresponding columns in the lower panel indicate the overlaps
%%   %% ($Z_{j}^n = \bra{0}O_j\ket{n}$) of each interpolating operator onto the finite-volume
%%   %% Hamiltonian eigenstates. Ratio fits to those levels below $K\pi\pi$ threshold are
%%   %% used in the final analysis.
%% \end{figure}

%% \begin{figure}[!ht]
%%   \centering
%%   \includegraphics{figures/summPlot_mr_mpi.tikz}
%%   \hspace{-0.5em}%
%%   \includegraphics{figures/summPlot_g_mpi.tikz}
%%   \caption{Summary of lattice QCD calculations of $K^{*}(892)$
%%     resonance parameters, together with phenomenological values (shown as asterisks)
%%     from~\cite{pdg2016} where the neutral values for the mass and width are
%%     taken. This choice gives consistent values to~\cite{Bali2016}, while
%%     hadro-produced $K^{*}(892)$ parameters result in a coupling which is about $5\%$
%%     larger. The statistical and systematic errors from~\cite{Wilson2015a} are added
%%     in quadrature.}
%%   \label{fig:kst}
%% \end{figure}

\section{Results: excited meson spectroscopy}

On two anisotropic ensembles $(32^3|230)$, $(24^3|390)$, with $m_\pi \sim 230, 390$ MeV
respectively, we consider all isovector, non-strange, bosonic channels with negative
parity and positive G-parity. In the interest of brevity, preliminary results are
presented here for the resonance-rich $T_{1u}^+$ channel in which spin-1 and spin-3
states are expected to appear. In this symmetry channel we use a carefully selected set
of 73 operators, 9 of which are so-called \textit{optimised} single-hadron (SH) operators
resultant from a preliminary GEVP using only SH operators chosen with care to best
produce the expected spin-1 and spin-3 states in this channel. The remaining 64
two-hadron (MH) operators are chosen guided by the spectrum of all possible two-hadron
states in the $T_{1u}^+$ symmetry channel assuming no interactions between the particles.
Optimal interpolators for the expected hadronic states are chosen based on preliminary
low-statistics runs and are described in detail in Ref.~\cite{Morningstar2013}.

\begin{figure}[thb]
  \centering
  \includegraphics[scale=0.7]{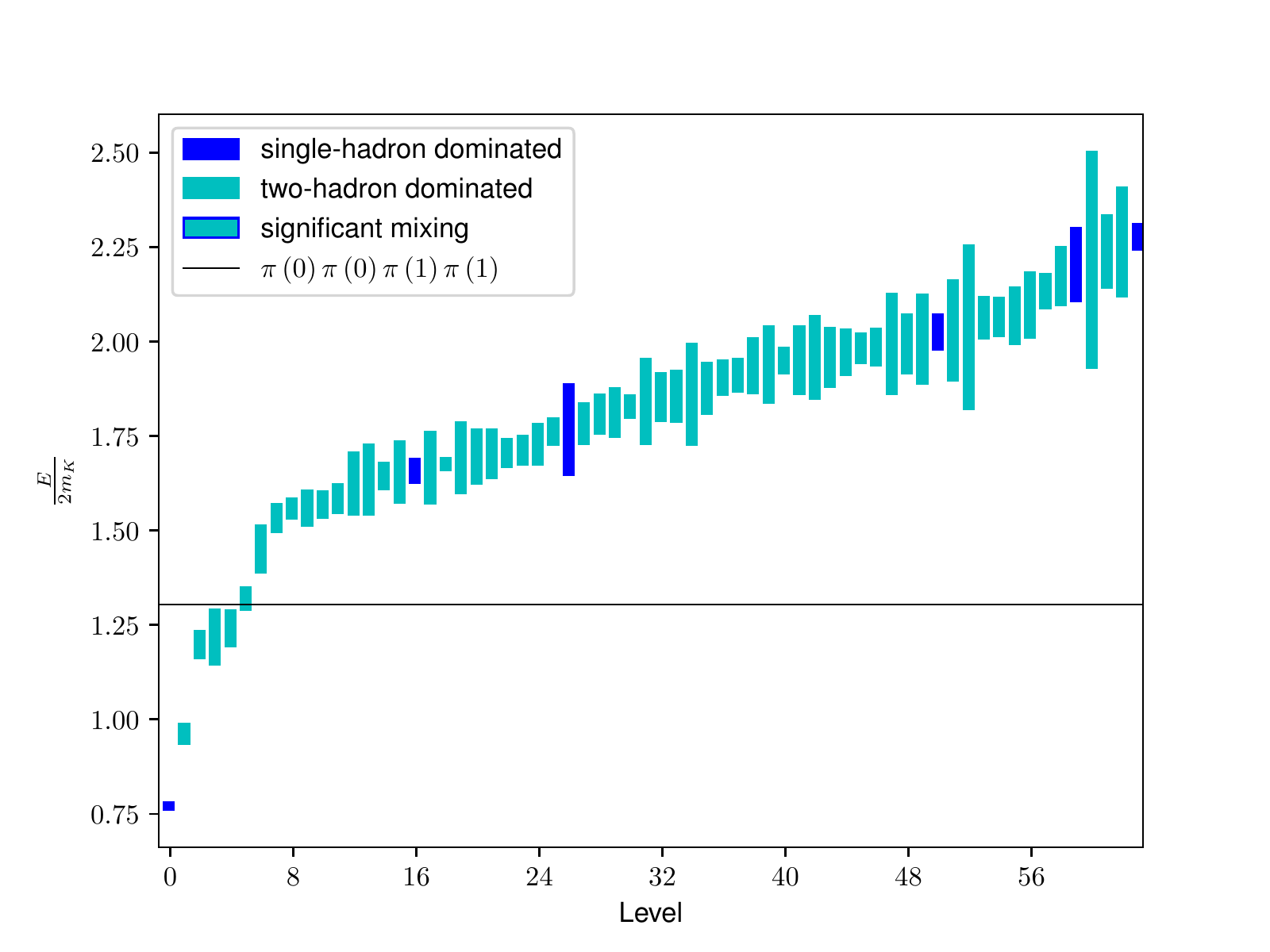}
  \caption{First 63 excited states in the isovector, non-strange $T_{1u}^+$ channel with
    $m_\pi \sim 230$ MeV. Levels with maximal overlap onto the optimised single hadron
    (SH) and two-hadron operators (MH) are indicated by solid blue and cyan boxes
    respectively. Spectrum extracted using multi-exponential fits to diagonal elements of
    Eq.~(\protect\ref{eq:rotated_corr}) from a GEVP with 73 operators (9 SH + 64 MH).
  \label{fig:spec_all}}
\end{figure}

\begin{figure}[bht]
  \centering
  \includegraphics[width=0.85\textwidth]{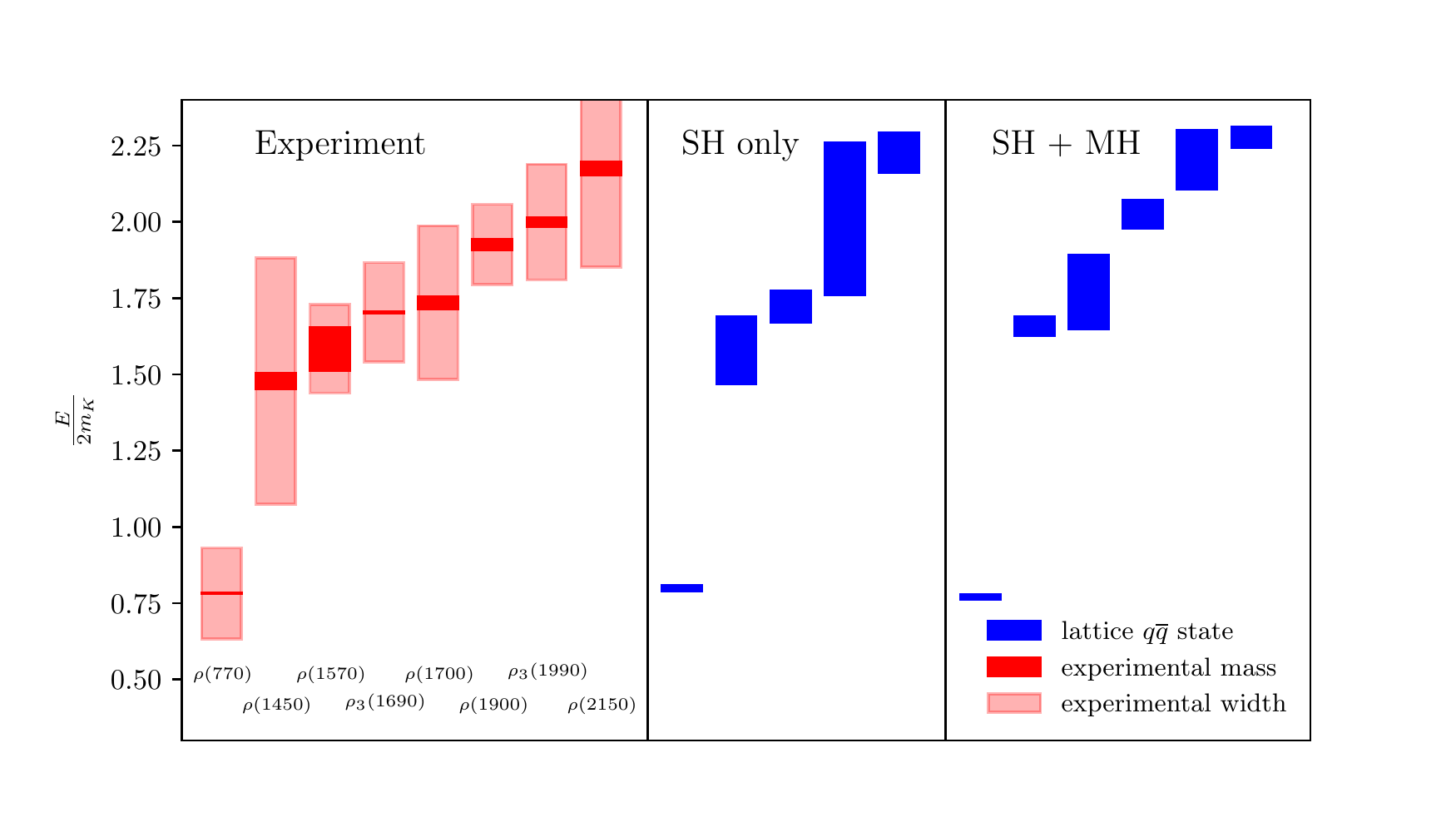}
  \caption{Comparison of experimental resonances (from Ref.~\cite{pdg2016}) to levels in
    the isovector, non-strange $T_{1u}^+$ channel with $m_\pi \sim 230$ MeV which have
    dominant overlap with the single-hadron interpolators. The central plot contains
    the spectrum $\lesssim 5 m_K$ as determined using only the single-hadron (SH)
    operators whereas the right column comes from the full spectrum determination using
    single- and multi-hadron (MH) operators. Note the emergence of an additional SH
    dominated level in this energy range with the inclusion of MH operators.
  \label{fig:expt_comp}}
\end{figure}

Using the $Z$ overlaps for each operator onto the finite-volume energy levels, level
identification can be performed based on the structure of the judiciously chosen probe
operators. QCD being a complicated interacting quantum field theory makes this
characterisation of stationary states difficult and largely qualitative. Nevertheless,
we are well able to identify the stationary states expected to evolve into the
single-meson resonances that correspond to quark-antiquark excitations in infinite-volume.
For $m_\pi \sim 230$ MeV a staircase plot summarising the $T_{1u}^+$ finite-volume
spectrum is shown in Fig.~\ref{fig:spec_all}. In Fig.~\ref{fig:expt_comp} we see that
many more resonant states are seen in experiment than in our finite-volume calculation.
There are multiple possible explanations for this, for example some states from
experiment may not be simple quark-antiquark excitations but more exotic molecular
states that would not be clearly identified by our SH probe operators. More interesting
however is the number of quark-antiquark identified stationary states found when
considering only SH operators versus when mixing with MH operators is included. The
appearance of an additional state in the energy range shown suggests that mixing between
quark-antiquark excitations and two-hadron states is crucial for the reproduction of some
states seen in experiment. Further investigation is required, with a finite-volume
Hamiltonian based analysis for qualitatively describing the stationary state spectrum in
progress. Note that we have not included any three- or four-particle interpolators
despite going above the thresholds for creating such states.

%% \section{Conclusions}
\acknowledgments
This work was supported by the U.S.~National Science Foundation
under award PHY-1613449.  Computing resources were provided by
the Extreme Science and Engineering Discovery Environment (XSEDE)
under grant number TG-MCA07S017.  XSEDE is supported by National
Science Foundation grant number ACI-1548562.

%% \clearpage
%\bibliography{refs}
%\bibliographystyle{JHEP}

\end{document}